\documentclass[a4paper]{article}
\usepackage{INTERSPEECH2021}
\usepackage[table]{xcolor}
\usepackage{amsmath,graphicx}
\usepackage{pdfpages}
\usepackage{hyperref}
\usepackage{booktabs}
\usepackage{multirow}
\usepackage{arydshln}
\usepackage{pifont}
\usepackage{url}
\usepackage{subcaption}
\usepackage{caption}
\definecolor{aliceblue}{rgb}{0.94, 0.97, 1.0}
\definecolor{mistyrose}{rgb}{1.0, 0.49, 0.88}
 
\def\prepara{{\vspace{5pt}}}

\newcommand{\new}[1]{\textcolor{black}{#1}}

\title{VoxSRC 2021: The Third VoxCeleb Speaker Recognition Challenge}

\name{Andrew Brown$^1$, Jaesung Huh$^{1}$, Joon Son Chung$^{1,2}$, Arsha Nagrani$^{1}$\footnotemark[2], \\ Daniel Garcia-Romero$^{3}$, Andrew Zisserman$^{1}$}

\address{$^1$Visual Geometry Group, Department of Engineering Science, University of Oxford, UK\\
$^2$Naver Corporation, South Korea \\
$^3$Amazon AWS AI, USA 
}
\email{\url{http://www.robots.ox.ac.uk/\~vgg/data/voxceleb/competition2021.html}}

\begin{document}
\maketitle
\begin{abstract}
The third instalment of the VoxCeleb Speaker Recognition Challenge was held in conjunction with Interspeech 2021. The aim of this challenge was to assess how well current speaker recognition technology is able to diarise and recognise speakers in unconstrained or `in the wild' data. The challenge consisted of: (i) the provision of publicly available speaker recognition and diarisation data from YouTube videos together with ground truth annotation and standardised evaluation software;  and (ii) a virtual public challenge and workshop held at Interspeech 2021.
This paper outlines the challenge, and describes the baselines, methods and results. We conclude with a discussion on the new multi-lingual focus of VoxSRC 2021, and on the progression of the challenge since the previous two editions.
\end{abstract}

\renewcommand*{\thefootnote}{\fnsymbol{footnote}}
\footnotetext[2]{Also at Google Research.}


\section{Introduction}
\label{sec:intro}
\new{In 2021, we held the third installment of the annual VoxCeleb Speaker Recognition Challenge~\cite{chung2019voxsrc,nagrani2020voxsrc} (VoxSRC). The primary goals of the VoxSRC speaker recognition challenges are to:} (i) explore and promote new research in speaker recognition ‘in the wild’; (ii) measure and calibrate the performance of the current state of technology through public evaluation tools; and (iii) provide open-source data freely accessible to all in the research community.

\new{In the second edition of the challenge~\cite{nagrani2020voxsrc}, we introduced a new self-supervised speaker verification track, as well as a diarisation track. New metrics were also introduced. For the third edition, we kept the tracks and metrics the same, and instead added a multi-lingual focus to the speaker verification tracks. The reason for this was twofold:  first, to promote the fairness and accessibility of speaker verification models, so as to allow people from diverse language groups to use these deep learning models; and  second, to provide a more challenging test set for the speaker verification tracks, given the near saturation of performance in previous years.}

In this paper, we describe the details of the evaluation task, the datasets provided, the challenge evaluation results and subsequent discussion. Further details can be found at the challenge website\footnote{\url{http://www.robots.ox.ac.uk/~vgg/data/voxceleb/competition2021.html}}. 

\section{Task Description} 
There were two tasks in this challenge, \textit{speaker verification} and \textit{speaker diarisation}. \textit{Speaker verification} is the task of determining whether a given pair of speech utterances are from the same speaker or not, while \textit{speaker diarisation} aims to break up multi-speaker audio into homogeneous single speaker segments, effectively solving ‘who spoke when’. Within the task of speaker verification we had three different tracks, each constraining the data allowed for training models, though with a common test and evaluation metrics. 

\subsection{Tracks} \label{sec:tracks}
The challenge consisted of the following four tracks:
\begin{enumerate}
    \item Speaker Verification -- Closed
    \item Speaker Verification -- Open
    \item Speaker Verification -- Self-supervised (Closed)
    \item Speaker diarisation -- Open
\end{enumerate}
\new{The tracks are identical in form to those in VoxSRC 2020~\cite{nagrani2020voxsrc}, although the test data for VoxSRC 2021 was new. For the verification tracks, the open and closed training conditions refer to the training data allowed, and are described in Sec.~\ref{sec:data}. For Track 3, participants \textit{could not} use any speaker labels during training, however they were allowed to use the visual modality (faces) as well from the videos.}
 



\subsection{Data}\label{sec:data}
The VoxCeleb datasets were still the primary datasets for the speaker recognition (Tracks 1--3). \new{For VoxSRC 2021, we added a multi-lingual focus to the the speaker verification tracks, in order to create more challenging validation and test sets, and also to promote/reward fairness and accessibility of speaker verification models. In VoxSRC 2020, \textbf{VoxConverse}~\cite{Chung20} was used for the speaker diarisation track. This year, we similarly use VoxConverse. As promised in 2020, we publicly released the VoxSRC 2020 test set and annotations for Track 4. This in turn became the validation set for 2021, and we introduced a new replacement test set from the same domain.} 
\subsubsection{Speaker Verification -- Track 1, 2 and 3}
The VoxCeleb datasets~\cite{nagrani2020voxceleb,Chung18a,Nagrani17} consist of speech segments from unconstrained YouTube videos for several thousand individuals, and were created using an automatic pipeline.
For a full description of the pipeline and an overview of the datasets, see~\cite{nagrani2020voxceleb}. 


\prepara\noindent\textbf{Train set (Closed and Open Conditions):} The closed training condition required that participants train only on the VoxCeleb2 dev dataset~\cite{Chung18a}, which contains 1,092,009 utterances from 5,994 speakers. For the open training condition, participants could use the VoxCeleb datasets and any other data, except for the challenge's \textit{test} data. 

\prepara\noindent\textbf{Validation and Test sets:} 
We provided a challenging validation set to participants to 
examine the performance of their models before uploading results to the evaluation server, in addition to the actual test set which was released a month before the challenge results were due. Unlike the validation set, the test set was \textit{blind}, i.e.\ the speech segments were released but with no annotations. The test data was released strictly for reporting of results alone, participants were not allowed to use this data in any way to train or tune systems.

The validation dataset consisted of trial pairs of speech from the identities in the VoxCeleb1 dataset, while the test set consisted of disjoint identities not present in either VoxCeleb1 or VoxCeleb2. Each trial pair consisted of two single-speaker audio segments, of variable length. \new{Last year, additional out-of-domain data was used in the form of speech segments from movies (VoxMovies~\cite{brown2020playing}, ) to make challenging validation and test sets. This year we did not use this data, but instead added a multi-lingual focus.}

\new{By the design of the dataset collection pipeline~\cite{nagrani2020voxceleb}, the VoxCeleb datasets consist of mainly English speaking speech segments. 
When forming the validation and test sets, a multi-lingual focus was added by putting an emphasis on sampling positive and negative pairs containing non-English speech segments. This requires language labels, which did not exist for the VoxCeleb datasets. We obtained the language labels using a three step pipeline consisting of a combination of automatic and manual annotation. First, we obtain automatic language predictions from a model trained on VoxLingua107~\cite{valk2021slt}. This model outputs the softmax predictions over 107 languages. Second, we manually annotated the correctness of the language predicted for each speech segment for the 12 most frequently occurring languages in VoxCeleb1 based on these predictions. Third, we used the manual annotations to obtain language-specific classification thresholds on the automatic predictions, and used the resulting thresholds to classify each of the speech segments in VoxCeleb1 as either one of these 12 languages or not.
The multi-lingual focus offers a significant challenge to state of the art speaker recognition systems (Section~\ref{method_and_results}).}

The statistics of the val and test sets can be found in Table~\ref{tab:testdata_track123}. The val and test data were checked using a combination of automatic and manual techniques for any errors using the same procedure described in \cite{nagrani2020voxceleb}, and following an identical procedure to VoxSRC19~\cite{chung2019voxsrc}.
Similarly to VoxSRC 2020, the challenge did not have same-session trials (e.g.\ segments from the same interview) in the test and validation sets. 

\subsubsection{Speaker Diarisation -- Track 4}
VoxConverse~\cite{Chung20} is a speaker diarisation dataset from diverse domains such as panel discussions, news segments and talk shows. It contains multi-speaker audio segments with challenging background conditions and overlapping speech. The dataset was constructed using a semi-automatic audio-visual pipeline, with a combination of active speaker detection~\cite{Chung16a}, source separation~\cite{Afouras18} and speaker verification~\cite{chung2020defence}. The dataset consists of a development, and test set. Please refer to \cite{Chung20} for more details.

\prepara\noindent\textbf{Training set:} Similar to last year, participants were allowed to use any data to train their models outside of the challenge's test set.

\prepara\noindent\textbf{Validation set:} In VoxSRC 2020, the VoxConverse test set was used as the hidden test set for Track 4, and the VoxConverse development set was used for validation. This year, as promised, we released the entire VoxConverse dataset (both development and test) for use in validation. Last year's test set contains 232 audio wavfiles, that were manually verified before release, and which can be downloaded from our website\footnote{\url{https://www.robots.ox.ac.uk/~vgg/data/voxconverse}}. Overall, the total duration of the VoxConverse set is approximately 64 hours. On average, the number of speakers per each audio segments is 4-6 and percentage of overlap speech is 3\%.

\prepara\noindent\textbf{Test set:} \new{We created a new test set for this challenge with the same semi-automatic pipeline used for VoxConverse. The outputs of the automatic pipeline using state-of-the-art audio-visual models were corrected by  annotators using the VIA annotation tool~\cite{dutta2019vgg}. The test set consists of 264 audio files with a total duration of 33 hours. Please refer to Table~\ref{table:testdata_track4} for detailed statistics of validation and test sets.}

\begin{table*}[h!]
    \centering
    \renewcommand\arraystretch{1.2}
    \small
    \begin{tabular}{ l  r  r  r  r r r r r}
    \toprule
     \textbf{set} & \textbf{\# audios} & \textbf{\# mins}  & \textbf{\# spks}  & \textbf{video durations (s)} & \textbf{speech \%} & \textbf{overlap \%} \\  
    \midrule 
     Track 4 val   & 448   & 3,830 & 1 / 5.5 / 21  & 22.0 / 512.9 / 1200.0     &    10.7 / 90.7 / 100.0   & 0 / 3.3 / 29.8 \\  
     Track 4 test   & 264   & 1,989  & 1 / 5.6 / 25 & 30.7 / 452.1 / 1200.0     &    14.2 / 91.5 / 99.3    & 0 / 3.0 / 37.1 \\ 
     \bottomrule
    \end{tabular} 
    \caption{\small{\new{Statistics of the speaker diarisation val and test sets (Track 4). 
    Entries that have 3 values are reported as min/mean/max. \textbf{\#~spks:} Number of speakers per video. \textbf{\#~mins:} Total duration of dataset in minutes. \textbf{video durations (s):} Length of videos in seconds. \textbf{speech \%:} Percentage of video time that is speech. \textbf{overlap \%:} Percentage of overlapping speech.}}}
    \label{table:testdata_track4}
\end{table*}

\begin{table}[]
    \centering
    
    \begin{tabular}{ccccc}
    \toprule
         & \textbf{\# Pairs} & \textbf{\# Utter.} & \textbf{Segment length (s)}  \\ \midrule
         val  & 60,000 & 64,711 & 3.96/8.11/144.92\\
         test & 476,224 & 116,984 & 2.04/5.01/81.04\\
         \bottomrule
    \end{tabular}
    \caption{\small{\new{Statistics of the speaker verification validation and test sets (Tracks 1--3). \textbf{\# Pairs} refers to the number of evaluation trial pairs, whereas \textbf{\# Utter.} refers to the total number of unique speech segments in the test set. Segment lengths are reported as min/mean/max.}}}
    \label{tab:testdata_track123}
\end{table}

\begin{table*}[]
    \renewcommand\arraystretch{1.1}
    \centering
    
    \begin{tabular}{cccccc}
    \toprule
         \textbf{Track} & \textbf{Rank} & \textbf{Team Name} & \textbf{Organisation} & \textbf{minDCF} & \textbf{EER}  \\ 
          \midrule
        \rowcolor{aliceblue} \multirow{4}{*}{1}  & - & Baseline & Provided & 0.351 & 5.88 \\
                  & 3 & JTBD~\cite{thienpondt2021idlab} & IDLab, Ghent University, Belgium & 0.129 & 2.27 \\ 
         & 2 & Beijing ZKJ-NPU~\cite{zhang2021beijing} & Beijing ZKJ Technology Ltd,  Northwestern Polytechnical Uni. & 0.118 & 2.84\\ 
        &1 & snowstar~\cite{zhao2021speakin} & SpeakIn Technologies Co. Ltd. & 0.103 & 1.85\\ 
         
         \hline
        \rowcolor{aliceblue} \multirow{4}{*}{1}  & - & Baseline & Provided & 0.351 & 5.88 \\
                  & 3 & JTBD~\cite{thienpondt2021idlab} & IDLab, Ghent University, Belgium & 0.131 & 2.05 \\ 
         & 2 & Beijing ZKJ-NPU~\cite{zhang2021beijing} & Beijing ZKJ Technology Ltd,  Northwestern Polytechnical Uni. & 0.118 & 2.84\\ 
        &1 & snowstar~\cite{zhao2021speakin} & SpeakIn Technologies Co. Ltd. & 0.103 & 1.85\\

         \midrule
        \rowcolor{aliceblue}\multirow{4}{*}{3}  & - & Baseline & Provided & 0.893 & 20.17 \\
         &3 & JaejinCho~\cite{cho2021jhu} & Johns Hopkins University, Baltimore, MD, USA  & 0.369 & 6.89\\
         &2 & phonexia~\cite{slavicek2021phonexia} & Phonexia Ltd. & 0.324 & 6.49\\ 
  &1 & DKU-DukeECE~\cite{cai2021dkudukeece} & Duke Kunshan University & 0.341 & 5.59\\ 
         \bottomrule
    \end{tabular}
    
    \caption{\small{Winners for the speaker verification tracks (Tracks 1, 2 and 3). For both metrics, a lower score
is better.}}
    \label{tab:results_verification}
\end{table*}

\begin{table*}[]
    \renewcommand\arraystretch{1.1}
    \centering
    \begin{tabular}{ccccc}
    \toprule
         \textbf{Rank} & \textbf{Team Name} & \textbf{Organisation} & \textbf{DER} & \textbf{JER}  \\ \midrule
\rowcolor{aliceblue}- & Baseline & Provided & 17.99 & 38.72 \\
         3 & njz~\cite{tencent2021} & Tencent AI Lab, China & 5.32 & 24.50\\ 
         2 & chen2101~\cite{wang2021bytedance} & Bytedance SAMI lab, China & 5.15 & 26.02\\ 
         1 & strato~\cite{wang2021dku} & Duke Kunshan Uni., China \& Duke Uni., USA \& Lenovo Research, China& 5.07 & 29.16\\ 
         \bottomrule
    \end{tabular}
    \caption{\small{Winners for the speaker diarisation track (Track 4). For both metrics, a lower score
is better. }}
    \label{tab:results_diarsiation}
\end{table*}

\section{Challenge Mechanics} 

\subsection{Evaluation metrics}
We released a validation toolkit\footnote{\url{https://github.com/JaesungHuh/VoxSRC2021}} for both speaker verification and speaker diarisation. Participants were encouraged to evaluate their models using this public code on the validation set of each track. \new{The evaluation metrics are the same as in VoxSRC 2020.}

\prepara\noindent\textbf{Speaker verification. }
For the speaker verification tracks (Tracks 1-3), we displayed two metrics, Equal Error Rate (EER) and minimum Detection Cost Function (minDCF). EER is a popular metric for evaluating the performance of speaker verification. It is used to determine the threshold value for a system when its false acceptance rate (FAR) and false rejection rate (FRR) are equal. minDCF ($C_{DET}$) can be computed as:
\begin{equation}
    C_{DET} = C_{miss} \times P_{miss} \times P_{tar} +C_{fa} \times P_{fa} \times (1 - P_{tar}) 
\label{eqn:dcf}
\end{equation}

This is same as the primary metric of the NIST SRE 2018 evaluation~\cite{nist2018}.
We set $C_{miss} = C_{fa} = 1$ and $P_{tar}=0.05$ in our cost function. 

For Tracks 1 and 2, the primary metric was minDCF and final ranking was determined by this score alone. For Track 3, the primary metric was EER. For both metrics, a lower score is better.

\prepara\noindent\textbf{Speaker diarisation.}
For Track 4, we adopted two diarisation metrics, Diarisation Error Rate (DER) and Jaccard Error Rate (JER). DER is used as a primary evaluation metric in this track.

DER is a standard evaluation metric for speaker diarisation. It is the sum of speaker error, false alarm speech and missed speech. We applied a forgiveness collar of 0.25 sec, and overlapping speech was not ignored.

We also reported the Jaccard error rate (JER), a metric introduced for the DIHARD II challenge that is based on the Jaccard index. The Jaccard index is a similarity measure typically used to evaluate the output of image segmentation systems and is defined as the ratio between the intersection and union of two segmentations. To compute Jaccard error rate, an optimal mapping between reference and system speakers is determined and for each pair the Jaccard index of their segmentations is computed. The Jaccard error rate is then 1 minus the average of these scores. For more details please consult Section 3 of the Dihard Challenge Report~\cite{ryant2019second}.


\subsection{Baselines} \label{baselines}
\new{Baseline methods were provided to all participants as a starting point for their development. This section describes these baselines for each of the challenge tracks}

\new{For Tracks 1 and 2, the baseline is adopted from the publicly released Naver Clova submission~\cite{heo2020clova} to the VoxSRC 2020 challenge~\cite{nagrani2020voxsrc}. The model is trained on 64-mel spectrograms with pre-emphasis as an input. The backbone architecture is the original ResNet-34~\cite{he2015deep} with attentive statistical pooling~\cite{okabe2018attentive}. A combination of angular prototypical loss~\cite{chung2020defence} and cross entropy loss is used as for the loss function. The network is only trained with VoxCeleb2 dev set and produces an EER of 1.18\% on VoxCeleb1 test set. This baseline achieved a minDCF of 0.351 and an EER of 5.88\% on VoxSRC 2021 challenge test set.}

\new{For the self-supervised track, the baseline model follows the popular contrastive approach~\cite{chen2020simple,falcon2020framework}, where different \textit{positive} views of the same instance are generated with diverse data augmentation techniques such as adding noise or reverberation. The backbone architecture is a Fast ResNet-34~\cite{chung2020defence} backbone.The detailed description can be found in~\cite{huh2020augmentation}. This model achieved a minDCF of 0.8925 and an EER of 20.17\%. This is the same baseline method as in VoxSRC 2020.}

\new{For the Track 4 baseline, we adopted a sliding-window approach where speaker identity is determined by extracting and clustering speaker embeddings (for more details, see~\cite{wang2018speaker}). A few changes were made from~\cite{wang2018speaker}: (1) We replaced the speaker embedding model with the baseline model from Tracks 1 and 2 for more robust representations. (2) Instead of the i-vector based GMM model introduced in the original paper, we used py-webrtcvad~\cite{pywebrtc} which is publicly available. (3) We used agglomerative hierarchical clustering (AHC) of speaker embeddings (this outperformed several other clustering algorithms). The resulting model achieves 17.99\% DER and 38.72\% JER on the challenge test set.} 

\subsection{Submission} 
\new{The challenge was hosted via CodaLab\footnote{\url{https://competitions.codalab.org/}}. We introduced two phases: ``Challenge workshop'' and ``Permanent'' and the challenge results were based on the former phase. Participants could only submit one submission per day and five submissions in total in order to avoid overfitting on the challenge test set. Submission for the ``Challenge workshop" phase was available until 1$^{st}$ of September, 2021.}
\new{Participants were required to submit reports of their submissions by 4$^{th}$  of September, 2021. The workshop was held on the 7$^{th}$ of September, 2021 in conjunction with Interspeech 2021.}

\section{Methods and Results} \label{method_and_results}

There were a total of 374 submission across all four tracks this year, an increase of 37\% compared to VoxSRC 2020. The top three ranked teams for each track are reported in Table~\ref{tab:results_verification}, along with their scores. The videos and slides of the winners' presentations are all available on our website \footnote{\url{http://www.robots.ox.ac.uk/~vgg/data/voxceleb/interspeech2021.html}}.

\prepara\noindent\textbf{Speaker verification.}
 The top three ranked teams were the same for both Tracks 1 and 2 (supervised tracks) this year, meaning that no additional training data outside of VoxCeleb was used for Track 2 submissions. The winning team~\cite{zhao2021speakin} utilised two state-of-the-art Convolutional Neural Network backbone architectures, RepVGG~\cite{ding2021repvgg} and ResNet~\cite{he2015deep}. Four variants of RepVGG architectures and three variants of ResNet architectures were explored along with two types of input: 81 and 96 log-Mel filterbanks. The team used strong augmentations, that intentionally went beyond the natural intra-class variation in the training set, to artificially create new classes. Specifically, they used three-fold speed augmentation~\cite{yamamoto2019speaker}, and perturbed the dataset by a factor of 0.9 or 1.1 based on the SoX speed function to artificially obtain a training set that was three times bigger in terms of number of classes than the original VoxCeleb2 dev set. 
 Diverse augmentation techniques, including gain augmentation, additive noise, room reverberation and time stretching were used to train the model to be more robust to challenging conditions. The training was designed with two stages. In the first stage, AM-Softmax loss~\cite{wang2018amsoftmax, wang2018cosface} with sub-centering method and inter-topK penalty is used as the training objective. Then, at the second stage, they adopted a large-margin fine-tuning (introduced in the VoxSRC 2020 challenge by~\cite{thienpondt2020idlab}) and replaced the loss function to AAM-softmax loss~\cite{deng2019arcface}. For post-processing, adaptive score normalisation~\cite{yin2008adaptive} using VoxCeleb2 dev set as cohorts and QMF rescoring~\cite{thienpondt2020idlab} are utilised to improve performance. The resulting performance were 0.103 minDCF and 1.85\% EER.
 
 The second placed team~\cite{zhang2021beijing} fused the predictions from several different models, including ECAPA-TDNN~\cite{desplanques2020ecapa}, several variants of ResNets, Res2Net~\cite{gao2019res2net} with squeeze-and-excitation module, D-TDNN-SE~\cite{yu2020densely} and CoAtNet~\cite{dai2021coatnet}. Matrix score averaging method and PLDA re-scoring were used for post-processing the predictions. This model produced a minDCF of 0.18 and an EER of 2.84\%. There was little difference between the first and second placed teams in terms of minDCF (the primary metric), but there was a significant 1\% difference in EER.
 
 In the self-supervised track, the first placed team~\cite{cai2021dkudukeece} extended their previous two-stage iterative labeling framework~\cite{cai2021iterative}, and for the first time in the VoxSRC challenges, leveraged both audio and visual data (this is permitted for the self-supervised track). Specifically, they leveraged the complementary information in the audio and visual streams to generate more robust pseudo-labels during the iterative labelling framework. The pseudo-labels from the different modalities were fused using a novel clustering ensemble technique. These fused labels were used to train both audio and visual embedding extractors to extract more robust speaker representations. 40-Mel spectrograms are used as input to ResNet-34 audio and visual encoders. The model produced 0.341 minDCF and 5.59\% of EER.  
 
 The second placed team~\cite{slavicek2021phonexia} explored various methods such as i-vector based rescoring~\cite{dehak2009support}, variational Bayesian PLDA~\cite{borgstrom2021unsupervised} and an approximate EM algorithm for PLDA training, but did not see significant benefit from these techniques. Their final submission is based on contrastive learning and iterative clustering method followed by adaptive ZT-norm score normalisation~\cite{matejka2017analysis}. This gave a final score of 0.324 minDCF and an EER of 6.49\%.
 
 \prepara\noindent\textbf{Multi-lingual aspect.}
 \new{As detailed in Section~\ref{sec:data}, the verification tracks had a multi-lingual focus this year, via the inclusion of more multi-lingual data in the VoxSRC 2021 test set. Here, we provide some analysis into the performance of the winning methods  from the supervised tracks (1st and 2nd place on Track 1) and our provided baseline (Section~\ref{baselines}) on this multi-lingual data.}

\begin{figure}[t]
        \includegraphics[width=0.45\textwidth]{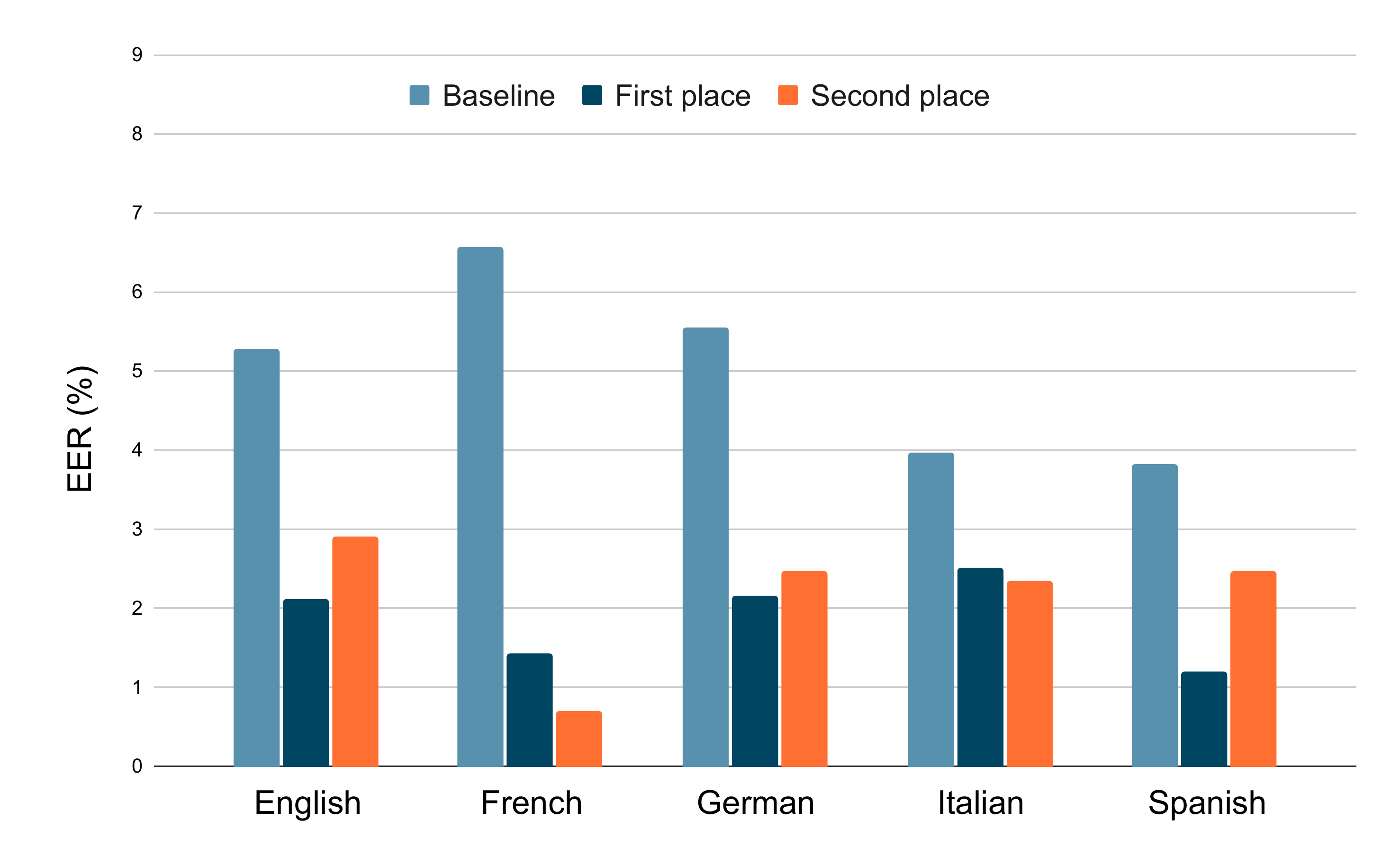}
          \caption{\small{\new{The performances of different models on the pairs only with certain languages in the VoxSRC 2021 test set. The metric is EER (lower is better)}}}
        \label{fig:withinlang}
\end{figure}

 \new{First, we analyse the ability of the methods to perform speaker verification \textit{within} different languages \textit{i.e.\ } when both speech segments in a trial verification pair come from the same language. We analyse this for the five most common languages in VoxCeleb. The results are shown in Figure~\ref{fig:withinlang}. We compute a per-language performance measure by computing the equal error rate from the same-language pairs. For all five  languages, there are at least 1000 same-language pairs (positive and negative) in the test set. The baseline model exhibits a lot of variance between the different languages. The winning methods both show far higher performance than the baseline across all languages, although there is still significant disparity in the performance between different languages. Interestingly, although English is the most common language in the training set, none of the models perform best on the English language pairs. However, we do note that some uncertainty via estimation error is introduced for the languages with a lower number of pairs \textit{e.g.\ } Spanish, relative to those with many pairs \textit{e.g.\ } English, French. This is due to the error introduced when estimating population statistics using small sample sizes. }
 
\begin{figure}[t]
        \includegraphics[width=0.45\textwidth]{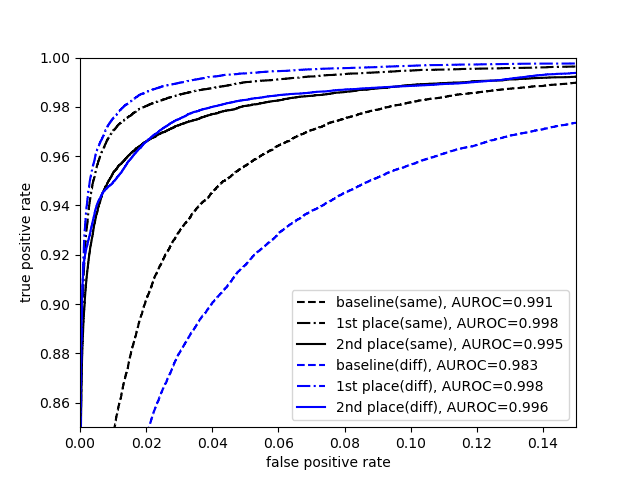}
          \caption{\small{\new{Receiver Operating Characteristic curves for different positive pairs. \textbf{1st place} and \textbf{2nd place} indicate the performance of top two winners on Track 1. \textbf{Same} is the subset of test set which positive pairs are always from same language while \textbf{diff} is the subset which positive pairs are always from different languages. Same negative pairs are used to plot both figures.}}}
        \label{fig:roc}
\end{figure}

 \new{Second, we analyse the model's ability to discriminate the same identity when speaking different languages, for example between speech segments of an actor providing interviews in both Spanish and French. These \textit{bi-lingual} samples present a significant challenge to verification methods due to their rarity in the training data. Figure~\ref{fig:roc} shows ROC curve plots on two different sets of positive pairs. One set contains positive pairs where both segments are the same language, and the other contains positive pairs where each segment is from a different language. The same negative pairs are used for both in order to be able to compare the two. The baseline model performs significantly worse on positive pairs with different languages, as demonstrated by the far lower area under the ROC. Surprisingly, the 1$^{st}$ and 2$^{nd}$ places of Track 1 perform similarly, or even slightly better in the different language pairs than same language pairs. This indicates that both models are relatively unaffected by the multi-lingual aspects of the hard positive pairs.}
 
 \new{Out of the winning teams in Tracks 1, 2, and 3, only one team~\cite{thienpondt2021idlab} used training methods specially designed for the multilingual focus. }
 
 \prepara\noindent\textbf{Speaker diarisation.}
 \new{Track 4 saw 106 submissions from 33 different teams this year. The performances of the top three ranked teams are shown in Table~\ref{tab:results_diarsiation}. Interestingly, the difference between the 1$^{st}$ place and the 10$^{th}$ place is less than 1\% on our primary metric, indicating the fierce competition this year, compared to when Track 4 was introduced in last year's challenge.}
 
 \new{The winner~\cite{wang2021dku} of this track employed a ResNet-LSTM based Voice activity detector (VAD), and trained a ResNet with global statistical pooling layer for a speaker embedding extractor. They explored two clustering methods: agglomerative hierarchical clustering (AHC) and LSTM-based spectral clustering~\cite{lin2019lstm}. For detecting overlapped regions, they investigated two overlapping models (using a similar architecture to the VAD), and a target-speaker voice activity detection model (TS-VAD). They re-arranged the challenge dev set by using the last 46 recordings as validation set and the rest for finetuning the model. The winner achieved 5.07\% DER on challenge test set.}
 
 \new{The second place~\cite{wang2021bytedance} team adopted a similar pipeline. They chose the \textit{pyannote 2.0}~\cite{bredin2020pyannote} voice activity detection module and ECAPA-TDNN speaker model~\cite{desplanques2020ecapa}. They adopted a two-stage clustering method, with an initial clustering stage using a variant of spectral clustering, and the re-clustering stage using AHC. The overlapped speech detection model introduced in ~\cite{landini2020analysis} was trained with DIHARD3~\cite{ryant2020third}, AMI corpus~\cite{carletta2007unleashing} and the VoxConverse dev set. The results from models trained with different hyper-parameters were fused with the DOVER-Lap~\cite{raj2021dover} algorithm. The resulting method showed a DER of 5.15\% on test set.}

 
\section{Workshop}
\new{Due to the ongoing COVID-19 pandemic, and in line with Interspeech 2021, the VoxSRC 2021 workshop was held entirely virtually as a Zoom webinar. Once again the workshop was free of cost for anybody to attend. The number of attendees peaked at 155 during the event, with a constant attendance of over 100 for the duration of the workshop. There were attendees from 30 different countries, spanning 5 continents (statistics gathered from the ticket distributor, EventBrite). The workshop consisted of an introductory talk from the organisers, announcements of the winners of each challenge track, short presentations from the winners where they gave an overview of their methods, and a keynote speech from Dr Andreas Stolke (Amazon Alexa Speech Organisation), titled ``Speaker Recognition and Diarisation for Alexa''. After each presentation from challenge winners, or the keynote, the speakers answered questions live from attendees. All slides and recorded videos from the workshop are available at our website\footnote{\url{http://www.robots.ox.ac.uk/~vgg/data/voxceleb/interspeech2021.html}}. The workshop was kindly sponsored by Naver Corporation, and AWS AI.}

 \begin{table*}[t!]
    \renewcommand\arraystretch{1.1}
    \centering
    \begin{tabular}{lccc}
    \toprule
         \textbf{Method} & \textbf{2019 test} & \textbf{2020 test}   & \textbf{2021 test}  \\ \midrule
         VoxSRC 2019 winner~\cite{zeinali2019but} & 1.42 & - & - \\
         \midrule
          VoxSRC 2020 winner ~\cite{thienpondt2020idlab} & 0.80 &  3.73 & -\\  
          VoxSRC 2020 2nd place~\cite{xiang2020xx205} & 0.75 & 3.81 & -\\
         \midrule
         VoxSRC 2021 winner~\cite{zhao2021speakin} & 0.57 & - & 1.85 \\  
          VoxSRC 2021 2nd place~\cite{zhang2021beijing} & 0.62 & - & 2.84 \\
          \bottomrule
    \end{tabular}
    \caption{\small{\new{Comparison of methods (\% EER) on the 2019, 2020, and 2021 test sets. The 2019 test set is contained in the test sets of 2020 and 2021, meaning performance can be compared via the 2019 test set. We compare the VoxSRC 2019 winning submission and the top-2 submissions from both VoxSRC 2020 and VoxSRC 2021 on the 2019 test set, showing the large performance improvement in a year. All results are shown on the closed track (Track 1). For \% EER shown, lower is better. }}}
    \label{tab:results_comparison}
\end{table*} 
\section{Related Challenges}
Track 1, 2 and 3 are focused on speaker recognition, which has been explored by the NIST-SRE (Speaker Recognition Evaluation) series~\cite{nist2018, sadjadi20172016, sadjadi20202019}, held since 1996 to measure state-of-the-art speaker recognition systems. Researchers from both academia and industry are encouraged to participate in NIST, however unlike NIST, all training data for VoxSRC is released publicly to the research community, even for those not participating in the challenge. Other challenges on speaker verification focus on noisy conditions~\cite{nandwana2019voices} or the far-field condition~\cite{qin2020interspeech}. 

Track 4 is complementary to several existing audio speaker diarisation challenges. The DIHARD challenges~\cite{ryant2018first,ryant2019second} are potentially the most popular. They evaluate state-of-the-art systems on extreme, ``hard'' conditions. Both the dev and test sets cover various background conditions, such as audiobooks, broadcast interviews, and restaurants. The third installment of the challenge~\cite{ryant2020third} will be concluded in early 2021.
Unlike VoxSRC, the challenge does not provide explicit training data, and hence any public or private data can be used for training models. Additionally, the DIHARD challenge applies \textit{no} forgiveness collar during evaluation and also has two separate diarisation tracks, one with oracle VAD and another with system VAD. 
Another popular challenge is the CHIME-6 challenge, where participants perform both speaker diarisation and speech recognition for multi-speaker conversations held in kitchen, dining and living room areas. The challenge data was made using binaural microphones and 4-channel microphone arrays, and the number of participants is fixed for each session. More details are provided at ~\cite{watanabe2020chime}.

\section{Discussion} \label{sec:discussion}
\new{The workshop had particularly high attendance this year, potentially due to the virtual format, but also reflecting the increased participation in the challenges from last year. Similarly to last year, all talks were pre-recorded and made accessible on the website, allowing for future access. We hope that in future the VoxSRC workshops will be in person, but to maintain this wide access, we will endeavor to keep recording and livestreaming presentations during future workshops. Participation in all four of the tracks increased significantly from last year (an increase of 21\%, 157\%, 140\%, and 94\% for Tracks 1, 2, 3, and 4 respectively, in terms of numbers of teams submitting results). }

\new{The significant increase in participation for Track 2 (157\%) is largely because this year, more teams from Track 1 submitted their Track 1 results into Track 2 as well. Interestingly, this year, the winning two teams for Track 2 only used training data from VoxCeleb. This is somewhat surprising, and could indicate that methods using unlimited training data do not have a clear performance advantage over methods with a smaller amount of limited training data. Explanations for this could be the introduction of new data augmentation techniques from the winning methods~\cite{zhao2021speakin}, and the fact that research groups with access to very large training training datasets (e.g.\ from industrial labs) are not submitting to the VoxSRC challenges.}

\new{The increase in participation in Tracks 3, and 4 (140\%, and 94\%, respectively) indicates a greater interest in the speech community in the areas of self-supervised learning and diarisation.
Interestingly, this year saw the first use in VoxSRC challenges of both audio and visual information by the winners of the self-supervised track~\cite{wang2021dku}. The complementary information between the two paired modalities offers great benefit to self-supervised methods.}

\new{The VoxSRC 2021 verification test set (Tracks 1, 2 and 3) was made more challenging this year by adding a multi-lingual focus. Every year, we include the entire VoxSRC 2019 test set in the verification test set, allowing us to compare methods from the different editions of the challenge. Table~\ref{tab:results_comparison} shows the performance of the top-2 submissions from VoxSRC 2021, VoxSRC 2020, and the winning submission from 2019 test set on both the 2019 test set, and the test set from their respective year. By examining the VoxSRC 2019 test performance, it is clear that the top-2 submissions from this year significantly outperform the winning submissions from the previous two years, demonstrating the vast improvement in speaker verification performance over one year.}

\new{Experiments with our verification baseline on the multi-lingual data introduced this year (Section~\ref{method_and_results}) showed that verification performance varies depending on language, and performs significantly worse on certain languages. The winning methods from Track 1 improved performance across all languages (Figure~\ref{fig:withinlang}), indicating that the field is moving in the right direction towards greater access to verification models, but there is still work to be done to provide equally high performance across all languages. Future work could involve increasing the amount of multi-modal data in the test set to reduce the uncertainty for less frequent languages.}

\new{We are unable to make a direct comparison between the 2020 and 2021 test sets, although the significantly higher \% EER of the 2020 test set from the 2020 winning method, over the \% EER of the 2020 test set from the 2020 winning method (3.73 vs 1.85) could indicate that the out-of-domain data used in 2020~\cite{brown2020playing} from movies~\cite{bain2020condensed} offers more of a challenge to speaker verification models than the multi-lingual focus of 2021.}

\section{Acknowledgements}
\new{This work is funded by the EPSRC programme grant EP/T028572/1 VisualAI}. Andrew Brown is funded by an EPSRC DTA Studentship. Jaesung Huh is funded by a Global Korea Scholarship. This work is also supported by Amazon AWS AI.
We also thank Rajan from Elancer and his team, \url{http://elancerits.com/}, for their huge assistance with diarisation annotation for VoxConverse and Bong-Jin Lee and his team in Naver Corporation for their effort to double-check the annotations. 


\clearpage
\raggedbottom
\bibliographystyle{IEEEtran}
\bibliography{shortstrings,refs}

\begin{thebibliography}{10}
\providecommand{\url}[1]{#1}
\csname url@samestyle\endcsname
\providecommand{\newblock}{\relax}
\providecommand{\bibinfo}[2]{#2}
\providecommand{\BIBentrySTDinterwordspacing}{\spaceskip=0pt\relax}
\providecommand{\BIBentryALTinterwordstretchfactor}{4}
\providecommand{\BIBentryALTinterwordspacing}{\spaceskip=\fontdimen2\font plus
\BIBentryALTinterwordstretchfactor\fontdimen3\font minus
  \fontdimen4\font\relax}
\providecommand{\BIBforeignlanguage}[2]{{%
\expandafter\ifx\csname l@#1\endcsname\relax
\typeout{** WARNING: IEEEtran.bst: No hyphenation pattern has been}%
\typeout{** loaded for the language `#1'. Using the pattern for}%
\typeout{** the default language instead.}%
\else
\language=\csname l@#1\endcsname
\fi
#2}}
\providecommand{\BIBdecl}{\relax}
\BIBdecl

\bibitem{chung2019voxsrc}
J.~S. Chung, A.~Nagrani, E.~Coto, W.~Xie, M.~McLaren, D.~A. Reynolds, and
  A.~Zisserman, ``Voxsrc 2019: The first voxceleb speaker recognition
  challenge,'' \emph{arXiv preprint arXiv:1912.02522}, 2019.

\bibitem{nagrani2020voxsrc}
A.~Nagrani, J.~S. Chung, J.~Huh, A.~Brown, E.~Coto, W.~Xie, M.~McLaren, D.~A.
  Reynolds, and A.~Zisserman, ``Voxsrc 2020: The second voxceleb speaker
  recognition challenge,'' \emph{arXiv preprint arXiv:2012.06867}, 2020.

\bibitem{Chung20}
J.~S. Chung, J.~Huh, A.~Nagrani, T.~Afouras, and A.~Zisserman, ``Spot the
  conversation: speaker diarisation in the wild,'' in \emph{Proc. Interspeech},
  2020.

\bibitem{nagrani2020voxceleb}
A.~Nagrani, J.~S. Chung, W.~Xie, and A.~Zisserman, ``Voxceleb: Large-scale
  speaker verification in the wild,'' \emph{Computer Speech \& Language},
  vol.~60, p. 101027, 2020.

\bibitem{Chung18a}
J.~S. Chung, A.~Nagrani, and A.~Zisserman, ``Voxceleb2: Deep speaker
  recognition,'' in \emph{Proc. Interspeech}, 2018.

\bibitem{Nagrani17}
A.~Nagrani, J.~S. Chung, and A.~Zisserman, ``{VoxCeleb:} a large-scale speaker
  identification dataset,'' in \emph{Proc. Interspeech}, 2017.

\bibitem{brown2020playing}
A.~Brown, J.~Huh, A.~Nagrani, J.~S. Chung, and A.~Zisserman, ``Playing a part:
  Speaker verification at the movies,'' \emph{arXiv preprint arXiv:2010.15716},
  2020.

\bibitem{valk2021slt}
J.~Valk and T.~Alum{\"a}e, ``{VoxLingua107}: a dataset for spoken language
  recognition,'' in \emph{Proc. IEEE SLT Workshop}, 2021.

\bibitem{Chung16a}
J.~S. Chung and A.~Zisserman, ``Out of time: automated lip sync in the wild,''
  in \emph{Workshop on Multi-view Lip-reading, ACCV}, 2016.

\bibitem{Afouras18}
T.~Afouras, J.~S. Chung, and A.~Zisserman, ``The conversation: Deep
  audio-visual speech enhancement,'' in \emph{Proc. Interspeech}, 2018.

\bibitem{chung2020defence}
J.~S. Chung, J.~Huh, S.~Mun, M.~Lee, H.~S. Heo, S.~Choe, C.~Ham, S.~Jung, B.-J.
  Lee, and I.~Han, ``In defence of metric learning for speaker recognition,''
  \emph{Proc. Interspeech}, 2020.

\bibitem{dutta2019vgg}
\BIBentryALTinterwordspacing
A.~Dutta and A.~Zisserman, ``The {VIA} annotation software for images, audio
  and video,'' in \emph{Proceedings of the 27th ACM International Conference on
  Multimedia}, ser. MM '19.\hskip 1em plus 0.5em minus 0.4em\relax New York,
  NY, USA: ACM, 2019. [Online]. Available:
  \url{https://doi.org/10.1145/3343031.3350535}
\BIBentrySTDinterwordspacing

\bibitem{thienpondt2021idlab}
J.~Thienpondt, B.~Desplanques, and K.~Demuynck, ``The idlab voxceleb speaker
  recognition challenge 2021 system description,'' 2021.

\bibitem{zhang2021beijing}
L.~Zhang, H.~Zhao, Q.~Meng, Y.~Chen, M.~Liu, and L.~Xie, ``Beijing zkj-npu
  speaker verification system for voxceleb speaker recognition challenge
  2021,'' 2021.

\bibitem{zhao2021speakin}
M.~Zhao, Y.~Ma, M.~Liu, and M.~Xu, ``The speakin system for voxceleb speaker
  recognition challange 2021,'' 2021.

\bibitem{cho2021jhu}
J.~Cho, J.~Villalba, and N.~Dehak, ``The jhu submission to voxsrc-21: Track
  3,'' 2021.

\bibitem{slavicek2021phonexia}
J.~Slavíček, A.~Swart, M.~Klčo, and N.~Brümmer, ``The phonexia voxceleb
  speaker recognition challenge 2021 system description,'' 2021.

\bibitem{cai2021dkudukeece}
D.~Cai and M.~Li, ``The dku-dukeece system for the self-supervision speaker
  verification task of the 2021 voxceleb speaker recognition challenge,'' 2021.

\bibitem{tencent2021}
N.~Zheng, N.~Li, Y.~Zhao, C.~Weng, and D.~Su, ``Tencent speaker diarization
  system for the voxceleb speaker recognition challenge 2021,''
  \url{https://www.robots.ox.ac.uk/~vgg/data/voxceleb/data_workshop_2021/reports/Tencent_diarization.pdf},
  2021.

\bibitem{wang2021bytedance}
K.~Wang, X.~Mao, H.~Wu, C.~Ding, C.~Shang, R.~Xia, and Y.~Wang, ``The bytedance
  speaker diarization system for the voxceleb speaker recognition challenge
  2021,'' \emph{arXiv preprint arXiv:2109.02047}, 2021.

\bibitem{wang2021dku}
W.~Wang, D.~Cai, Q.~Lin, L.~Yang, J.~Wang, J.~Wang, and M.~Li, ``The
  dku-dukeece-lenovo system for the diarization task of the 2021 voxceleb
  speaker recognition challenge,'' \emph{arXiv preprint arXiv:2109.02002},
  2021.

\bibitem{nist2018}
\emph{NIST 2018 Speaker Recognition Evaluation Plan}, 2018 (accessed 31 July
  2020),
  \url{https://www.nist.gov/system/files/documents/2018/08/17/sre18_eval_plan_2018-05-31_v6.pdf},
  See Section 3.1.

\bibitem{ryant2019second}
N.~Ryant, K.~Church, C.~Cieri, A.~Cristia, J.~Du, S.~Ganapathy, and
  M.~Liberman, ``The second dihard diarization challenge: Dataset, task, and
  baselines,'' \emph{arXiv preprint arXiv:1906.07839}, 2019.

\bibitem{heo2020clova}
H.~S. Heo, B.-J. Lee, J.~Huh, and J.~S. Chung, ``Clova baseline system for the
  {VoxCeleb} speaker recognition challenge 2020,'' \emph{arXiv preprint
  arXiv:2009.14153}, 2020.

\bibitem{he2015deep}
K.~He, X.~Zhang, S.~Ren, and J.~Sun, ``Deep residual learning for image
  recognition,'' in \emph{Proc. CVPR}, 2016.

\bibitem{okabe2018attentive}
K.~Okabe, T.~Koshinaka, and K.~Shinoda, ``Attentive statistics pooling for deep
  speaker embedding,'' \emph{arXiv preprint arXiv:1803.10963, 2018}, 2018.

\bibitem{chen2020simple}
T.~Chen, S.~Kornblith, M.~Norouzi, and G.~Hinton, ``A simple framework for
  contrastive learning of visual representations,'' \emph{arXiv preprint
  arXiv:2002.05709}, 2020.

\bibitem{falcon2020framework}
W.~Falcon and K.~Cho, ``A framework for contrastive self-supervised learning
  and designing a new approach,'' \emph{arXiv preprint arXiv:2009.00104}, 2020.

\bibitem{huh2020augmentation}
J.~Huh, H.~S. Heo, J.~Kang, S.~Watanabe, and J.~S. Chung, ``Augmentation
  adversarial training for unsupervised speaker recognition,'' in
  \emph{Workshop on Self-Supervised Learning for Speech and Audio Processing,
  NeurIPS}, 2020.

\bibitem{wang2018speaker}
Q.~Wang, C.~Downey, L.~Wan, P.~A. Mansfield, and I.~L. Moreno, ``Speaker
  diarization with lstm,'' in \emph{Proc. ICASSP}.\hskip 1em plus 0.5em minus
  0.4em\relax IEEE, 2018, pp. 5239--5243.

\bibitem{pywebrtc}
``Webrtc voice activity detector,'' 2021 (accessed 31 May 2021),
  \url{https://github.com/wiseman/py-webrtcvad}.

\bibitem{ding2021repvgg}
X.~Ding, X.~Zhang, N.~Ma, J.~Han, G.~Ding, and J.~Sun, ``Repvgg: Making
  vgg-style convnets great again,'' in \emph{Proceedings of the IEEE/CVF
  Conference on Computer Vision and Pattern Recognition}, 2021, pp.
  13\,733--13\,742.

\bibitem{yamamoto2019speaker}
H.~Yamamoto, K.~A. Lee, K.~Okabe, and T.~Koshinaka, ``Speaker augmentation and
  bandwidth extension for deep speaker embedding.'' in \emph{Proc.
  Interspeech}, 2019, pp. 406--410.

\bibitem{wang2018amsoftmax}
F.~Wang, W.~Liu, H.~Liu, and J.~Cheng, ``Additive margin softmax for face
  verification,'' \emph{arXiv preprint arXiv:1801.05599}, 2018.

\bibitem{wang2018cosface}
H.~Wang, Y.~Wang, Z.~Zhou, X.~Ji, D.~Gong, J.~Zhou, Z.~Li, and W.~Liu,
  ``Cosface: Large margin cosine loss for deep face recognition,'' in
  \emph{Proceedings of the IEEE conference on computer vision and pattern
  recognition}, 2018, pp. 5265--5274.

\bibitem{thienpondt2020idlab}
J.~Thienpondt, B.~Desplanques, and K.~Demuynck, ``The {IDLAB} voxceleb speaker
  recognition challenge 2020 system description,'' \emph{arXiv preprint
  arXiv:2010.12468}, 2020.

\bibitem{deng2019arcface}
J.~Deng, J.~Guo, N.~Xue, and S.~Zafeiriou, ``Arcface: Additive angular margin
  loss for deep face recognition,'' in \emph{Proc. CVPR}, 2019.

\bibitem{yin2008adaptive}
S.-C. Yin, R.~Rose, and P.~Kenny, ``Adaptive score normalization for
  progressive model adaptation in text independent speaker verification,'' in
  \emph{Proc. ICASSP}.\hskip 1em plus 0.5em minus 0.4em\relax IEEE, 2008, pp.
  4857--4860.

\bibitem{desplanques2020ecapa}
B.~Desplanques, J.~Thienpondt, and K.~Demuynck, ``Ecapa-tdnn: Emphasized
  channel attention, propagation and aggregation in tdnn based speaker
  verification,'' \emph{Proc. Interspeech}, 2020.

\bibitem{gao2019res2net}
S.~Gao, M.-M. Cheng, K.~Zhao, X.-Y. Zhang, M.-H. Yang, and P.~H. Torr,
  ``Res2net: A new multi-scale backbone architecture,'' \emph{IEEE transactions
  on pattern analysis and machine intelligence}, 2019.

\bibitem{yu2020densely}
Y.-Q. Yu and W.-J. Li, ``Densely connected time delay neural network for
  speaker verification.'' in \emph{Proc. Interspeech}, 2020, pp. 921--925.

\bibitem{dai2021coatnet}
Z.~Dai, H.~Liu, Q.~V. Le, and M.~Tan, ``Coatnet: Marrying convolution and
  attention for all data sizes,'' \emph{arXiv preprint arXiv:2106.04803}, 2021.

\bibitem{cai2021iterative}
D.~Cai, W.~Wang, and M.~Li, ``An iterative framework for self-supervised deep
  speaker representation learning,'' in \emph{Proc. ICASSP}.\hskip 1em plus
  0.5em minus 0.4em\relax IEEE, 2021, pp. 6728--6732.

\bibitem{dehak2009support}
N.~Dehak, R.~Dehak, P.~Kenny, N.~Br{\"u}mmer, P.~Ouellet, and P.~Dumouchel,
  ``Support vector machines versus fast scoring in the low-dimensional total
  variability space for speaker verification,'' in \emph{Proc. Interspeech},
  2009.

\bibitem{borgstrom2021unsupervised}
B.~J. Borgstr{\"o}m, ``Unsupervised bayesian adaptation of plda for speaker
  verification,'' in \emph{Proc. Interspeech}, 2021, pp. 1039--1043.

\bibitem{matejka2017analysis}
P.~Matejka, O.~Novotn{\`y}, O.~Plchot, L.~Burget, M.~D. S{\'a}nchez, and
  J.~Cernock{\`y}, ``Analysis of score normalization in multilingual speaker
  recognition.'' in \emph{Proc. Interspeech}, 2017, pp. 1567--1571.

\bibitem{lin2019lstm}
Q.~Lin, R.~Yin, M.~Li, H.~Bredin, and C.~Barras, ``Lstm based similarity
  measurement with spectral clustering for speaker diarization,'' \emph{arXiv
  preprint arXiv:1907.10393}, 2019.

\bibitem{bredin2020pyannote}
H.~Bredin, R.~Yin, J.~M. Coria, G.~Gelly, P.~Korshunov, M.~Lavechin, D.~Fustes,
  H.~Titeux, W.~Bouaziz, and M.-P. Gill, ``Pyannote. audio: neural building
  blocks for speaker diarization,'' in \emph{Proc. ICASSP}.\hskip 1em plus
  0.5em minus 0.4em\relax IEEE, 2020, pp. 7124--7128.

\bibitem{landini2020analysis}
F.~Landini, O.~Glembek, P.~Mat{\v{e}}jka, J.~Rohdin, L.~Burget, M.~Diez, and
  A.~Silnova, ``Analysis of the but diarization system for voxconverse
  challenge,'' \emph{arXiv preprint arXiv:2010.11718}, 2020.

\bibitem{ryant2020third}
N.~Ryant, K.~Church, C.~Cieri, J.~Du, S.~Ganapathy, and M.~Liberman, ``Third
  dihard challenge evaluation plan,'' \emph{arXiv preprint arXiv:2006.05815},
  2020.

\bibitem{carletta2007unleashing}
J.~Carletta, ``Unleashing the killer corpus: experiences in creating the
  multi-everything ami meeting corpus,'' \emph{Language Resources and
  Evaluation}, vol.~41, no.~2, pp. 181--190, 2007.

\bibitem{raj2021dover}
D.~Raj, L.~P. Garcia-Perera, Z.~Huang, S.~Watanabe, D.~Povey, A.~Stolcke, and
  S.~Khudanpur, ``Dover-lap: A method for combining overlap-aware diarization
  outputs,'' in \emph{IEEE Spoken Language Technology Workshop}.\hskip 1em plus
  0.5em minus 0.4em\relax IEEE, 2021, pp. 881--888.

\bibitem{zeinali2019but}
H.~Zeinali, S.~Wang, A.~Silnova, P.~Mat{\v{e}}jka, and O.~Plchot, ``{BUT}
  system description to voxceleb speaker recognition challenge 2019,''
  \emph{arXiv preprint arXiv:1910.12592}, 2019.

\bibitem{xiang2020xx205}
X.~Xiang, ``The xx205 system for the voxceleb speaker recognition challenge
  2020,'' \emph{arXiv preprint arXiv:2011.00200}, 2020.

\bibitem{sadjadi20172016}
S.~O. Sadjadi, T.~Kheyrkhah, A.~Tong, C.~S. Greenberg, D.~A. Reynolds,
  E.~Singer, L.~P. Mason, and J.~Hernandez-Cordero, ``The 2016 nist speaker
  recognition evaluation.'' in \emph{Proc. Interspeech}, 2017, pp. 1353--1357.

\bibitem{sadjadi20202019}
S.~O. Sadjadi, C.~Greenberg, E.~Singer, D.~Reynolds, L.~Mason, and
  J.~Hernandez-Cordero, ``The 2019 nist speaker recognition evaluation cts
  challenge,'' \emph{Proc. Speaker Odyssey}, 2020.

\bibitem{nandwana2019voices}
M.~K. Nandwana, J.~Van~Hout, M.~McLaren, C.~Richey, A.~Lawson, and M.~A.
  Barrios, ``The voices from a distance challenge 2019 evaluation plan,''
  \emph{arXiv preprint arXiv:1902.10828}, 2019.

\bibitem{qin2020interspeech}
X.~Qin, M.~Li, H.~Bu, W.~Rao, R.~K. Das, S.~Narayanan, and H.~Li, ``The
  interspeech 2020 far-field speaker verification challenge,'' \emph{arXiv
  preprint arXiv:2005.08046}, 2020.

\bibitem{ryant2018first}
N.~Ryant, K.~Church, C.~Cieri, A.~Cristia, J.~Du, S.~Ganapathy, and
  M.~Liberman, ``First dihard challenge evaluation plan,'' \emph{2018, tech.
  Rep.}, 2018.

\bibitem{watanabe2020chime}
S.~Watanabe, M.~Mandel, J.~Barker, and E.~Vincent, ``Chime-6 challenge:
  Tackling multispeaker speech recognition for unsegmented recordings,''
  \emph{arXiv preprint arXiv:2004.09249}, 2020.

\bibitem{bain2020condensed}
M.~Bain, A.~Nagrani, A.~Brown, and A.~Zisserman, ``Condensed movies: Story
  based retrieval with contextual embeddings,'' \emph{ACCV}, 2020.

\end{thebibliography}


\end{document}